# Approximate Bound State Solutions of the Fractional Schrödinger Equation under the Spin-Spin-Dependent Cornell Potential

**M. Abu-Shady[a], E. Omugbe[b] and E.P. Inyang[c]**

[a]Department of Mathematics and Computer Science, Faculty of Science, Menoufia University, Egypt

[b]Department of Physics, University of Agriculture and Environmental Sciences, Umuagwo, P.M.B. 1038, Imo State, Nigeria

[c]Department of Physics, National Open University of Nigeria, Jabi, Abuja, Nigeria

**Abstract**

In this work, the approximate bound state solutions of the fractional Schrödinger equation under a spin-spin-dependent Cornell potential are obtained via the convectional Nikiforov-Uvarov approach. The energy spectra are applied to obtain the mass spectra of the heavy mesons such as bottomonium, charmonium and bottom-charm. The masses for the singlet and triplet spin numbers increase as the quantum numbers increase. The fractional Schrödinger equation improves the mass spectra compared to the masses obtained in the existing literature. The bottomonium masses agree with the experimental data of the Particle Data Group where percentage errors for fractional parameters of $\beta = 1, \alpha = 0.97$ and $\beta = 1, \alpha = 0.50$ were found to be 0.67% and 0.49% respectively. The respective percentage errors of 1.97% and 1.62% for fractional parameters of $\beta = 1, \alpha = 0.97$ and $\beta = 1, \alpha = 0.50$ were obtained for charmonium meson. The results indicate that the potential curves coupled with the fractional parameters account for the short-range gluon exchange between the quark-antiquark interactions and the linear confinement phenomena which is associated with the quantum chromo-dynamic and phenomenological potential models in particle and high-energy physics

**Keywords:** Spin interaction potential; Nikiforov-Uvarov method; Generalized fractional derivative; meson masses

1. **Introduction**

Particles and their interactions are the fundamental building blocks of nature and may be produced or annihilated in high-energy accelerators. The elementary particles interact via the four forces of nature such as the strong nuclear force, weak, electromagnetic and the gravitational force. Of these particles, the quarks are the smallest which mediate the strong force and their interaction can be understood with a theory built on quantum chromo-dynamics and phenomenology. The quarks are composite structure, spin $1/2$ fermions and possess anti-particles. Two quarks constitute a meson, while three quark pairs combine to form baryons. The mesons and baryons are further classed as hadrons. Owing to the vital roles quarks play in understanding the evolvement of the universe, many particle and high-energy physicists have contributed significant research to study their properties. The rate of decay and spin average masses are among the properties of hadrons studied where the mass spectra is directly related to the energy of the hadrons. The bound state solutions of the wave equation with a quark anti-quark interaction and a phenomenological potential function have been used as primary properties in determining other properties of hadrons. The nature of the potential consists of a linear confinement function and a short-range Coulomb potential which is responsible for the gluon exchange between the quark and its anti-particle. In contrast, molecular potential functions have been used to investigate the mass spectra [1,2]. Douglas et al. [1] used the rovibrational model with the molecular Morse potential to obtain the resonances and masses of hadrons. Their findings were in strong accord with those of experiments and theoretical research. With the addition of the spin-spin, spin-orbit, and tensor components, Fang et al. [2] utilized the rovibrational model to derive the s-wave mass splitting between the singlet and triplet spin states as well as the triplet spin states for orbital quantum number larger than zero. The quarkonia energy spectra is a crucial source of knowledge on the quarks interaction [3].

The Schrödinger equation (SE) could be used to describe many aspects of heavy quarkonia, such as mass spectra (MS), and decay properties. It is generally accepted that the non-relativistic method is suitable in describing the heavy quarks and the level spacing between the energy levels is less than the component masses [4,5]. Different types of the inter-quark potential have been applied in heavy quarkonia mass spectroscopy [6,7]. Energy-dependent potentials have been applied to study of properties of mesons within the context of the SE [8].

Hadronic particles like mesons are made up of the quarks and their antiparticles. It is possible to examine the properties of hadrons using potential models [9-13]. The use of the non-relativistic method delivers a substantial description of the characteristics of the heavy mesons, including their MS, rates of decay, radius, and other characteristics. We are thus driven by fundamental features of quantum chromo-dynamics (QCD), specifically the spin dependency of the strong interaction fields in this study. The investigation of the properties of mesons such as the decay constant and the radius are not possible to reach across the entire range of separations from the QCD principle. The spin components of the quark-antiquark potentials have often been neglected in the literature since it is difficult to obtain an exact solution [14–17]. In these circumstances, numerical solutions are widely applied [18–21]. Recent research has suggested that two different forms of quark interactions can represent different meson sectors [22]. Kher and Rai [23] were able to determine the mass splitting's and decays characteristics of the charmonium meson utilizing the spin-dependent Cornell potential (CP). Using the CP, Luz et al. [25] solved the SE in the phase space. The Wigner function and related Airy function of the charmonium meson were examined using their derived solutions. Additionally, Gupta and Mehrotra [26] evaluated the masses, decay rates, and position expectation values of heavy mesons utilizing the energy-dependent potential. Extensive works on the properties of the quarkonia systems can be found in the literature [27-37].

The motivation for this research is as follows: I. To obtain an analytical solution for the spin-dependent SE utilizing the CP and generalized fractional derivative. II. To apply the energy spectra(ES) to obtain the MS of the heavy mesons. To our knowledge, this research is original and hasn't been published anywhere else.

The outline of our paper is as follows. In section 2, we give a brief review of the generalized fractional derivative (GFD) and the generalized fractional Nikiforov-Uvarov (NU) method. In the section 3, we solve the SE of the quarkonia under spin-spin interaction. Results and discussion are presented in the section 4. Finally, a brief concluding remarks is given in the section 5.

**2- The Generalized Fractional Derivative**

The idea of fractional calculus has recently attracted a lot of interest in a number of physics fields that deal with complex nonlinear events [38–42]. In these studies, the conformable fractional derivative was used in conjunction with the NU approach to effectively solve the fractional SE. The generalized fractional derivative is a new concept for the fractional derivative proposed by Abu-Shady and Kaabar [43,44]. This new definition coincides with the classical fractional derivative and has become a useful tool for solving fractional differential equations. It has been successfully applied in heavy meson and molecular physics, as can be seen in the references [45-49]. The GFD is a novel formula for a fractional derivative. Theorems that have been satisfied in the GFD, which offers a new approach for solving fractional differential equations (see Ref. [44]), include the Rolle('s) theorem, the mean value theorem, the derivative of two functions, the derivative of the quotient of two functions, and others. It has been argued that these definitions have more benefits than other traditional Caputo and Riemann-Liouville fractional derivative definitions.. For a function $Z : (0,\infty) \to R$, the generalized fractional derivative of order $0 < \alpha \leq 1$ of $Z(t)$ at $> 0$ is defined as

$$D^{GFD}Z(t) = \lim_{\varepsilon \to 0} \frac{Z\left(t + \frac{\Gamma(\beta)}{\Gamma(\beta-\alpha+1)}\varepsilon t^{1-\alpha}\right) - Z(t)}{\varepsilon}; \beta > -1, \beta \in R^+ \tag{1}$$

The properties of the generalized fractional derivative are,

I.   $D^\alpha[Z_{nl}(t)] = k_1 \, t^{1-\alpha} \, \dot{Z}_{nl}(t),$ \hfill (2)

II.  $D^\alpha[D^\alpha Z(t)] = k_1^2 \, [(1-\alpha) \, t^{1-2\alpha} \, \dot{Z}_{nl}(t) + t^{2-2\alpha} \, Z_{nl}''(t)],$ \hfill (3)

where, $k_1 = \frac{\Gamma[\beta]}{\Gamma[\beta-\alpha+1]},$ with $0 < \alpha \leq 1, 0 < \beta \leq 1$

III. $D^\alpha D^\beta t^m = D^{\alpha+\beta} t^m$ for function derivative of $Z(t) = t^m$, $m \in R^+$

IV.  $D^{GFD}(XY) = X D^{GFD}(Y) + Y D^{GFD}(X)$ where $X, Y$ be $\alpha-$ differentiable function

V.   $D^{GFD}\left(\frac{X}{Y}\right) = \frac{Y D^{GFD}(X) - X D^{GFD}(Y)}{Y^2}$ where $X, Y$ be α- differentiable function

VI.  $D^\alpha I_\alpha Z(t) = Z(t)$ for $\geq 0$ and $Z$ is any continuous function in the domain.

**2.1 The Generalized Fractional Nikiforov-Uvarov (NU) Method**.

In the fractional form as in Reference [50], the second-order parametric generalized differential equation is precisely solved.

$$D^\alpha[D^\alpha \psi(s)] + \frac{\bar{\tau}(s)}{\sigma(s)} D^\alpha \psi(s) + \frac{\bar{\sigma}}{\sigma^2} \psi(s) = 0, \tag{4}$$

where $\bar{\sigma}(s), \sigma(s)$ and $\bar{\tau}(s)$ are polynomials of $2\alpha-$th, $2\alpha-$th and $\alpha-$th degree.

where,

$$\pi(s) = \frac{D^\alpha \sigma(s) - \bar{\tau}(s)}{2} \pm \sqrt{\left(\frac{D^\alpha \sigma(s) - \bar{\tau}(s)}{2}\right)^2 - \bar{\sigma}(s) + K \sigma(s)},$$

(5)

and

$$\lambda = K + D^\alpha \pi(s), \tag{6}$$

$\lambda$ is constant and $\pi(s)$ is $\alpha-$th degree polynomial. The values of $K$ in the square root of Eq. (5) is possible to determine whether the expression under the square root is square of expression. Replacing $K$ into Eq. (5), we define

$$\tau(s) = \bar{\tau}(s) + 2\pi(s), \tag{7}$$

the derivative of $\tau$ should be negative [51], since $\rho(s) > 0$ and $\sigma(s) > 0$ then this is solution. If $\lambda$ in Eq.(6) is

$$\lambda = \lambda_n = -n D^\alpha \tau - \frac{n(n-1)}{2} D^\alpha[D^\alpha \sigma(s)]. \tag{8}$$

The hypergeometric type equation has a particular solution with degree $\alpha$. Eq. (4) has a solution which is the product of two independent parts

$$\psi(s) = \phi(s) y(s), \tag{9}$$

where,

$$y_n(s) = \frac{B_n}{\rho(s)} (D^\alpha)^n (\sigma(s)^n \rho_n(s)), \tag{10}$$

So, we can determine $\rho(s)$ as follows

$$D^\alpha[\sigma(s)\rho(s)] = \tau(s)\sigma(s), \tag{11}$$

Also, we can determine $\phi(s)$ as follows

$$\frac{D^\alpha \phi(s)}{\phi(s)} = \frac{\pi(s)}{\sigma(s)} \tag{12}$$

## 3-The solutions of the energy and wave function of quarkonia under spin-dependent Cornell potential.

The interaction potential takes the form [52]

$$V(r) = \frac{-4\alpha_s}{3r} + br + \frac{16\alpha_s \pi \left(\frac{\sigma}{\sqrt{\pi}}\right)^3 e^{-\sigma^2 r^2}\left(s(s+1)-\frac{3}{2}\right)}{9m_q m_{\bar{q}}}, \tag{13}$$

where $b, \alpha_s$ and $\sigma$ are constants that will be solved later from the experimental data [53,54]. The notations $m_q, m_{\bar{q}}$ denote the masses of the quark and its anti-particle and $s$ represent the spin quantum number. The colour factor is $-4/3$ multiplying the first term in the potential. The quantum number $J^{PC}$, is a representation of the hyperfine splitting between the spin quantum states. The symbolisations $P((-1)^{L+1})$ and $C((-1)^{L+s})$ are the respective parity and charge conjugation. The quantum number J(L+s) is obtained by adding the orbital and spin components. The singlet($s = 0$) and triplet($s = 1$) states are derived using the states $n^{2s+1}L_J$. The SE with energy eigenvalue $(E_{nl})$, wave function $(R_{nl}(r))$, and the reduced mass $\mu$ takes the form

$$\frac{\hbar^2}{2\mu}\frac{d^2\psi}{dr^2} + V_{eff}(r)\psi_{nl}(r) = E_{nl}\psi_{nl}(r) \tag{14}$$

With the potential in (13) and the Gaussian function present, Eq.(14) cannot be analytically solved. To solves Eq. (14), the Gaussian function is approximated by Taylor expanding it to second-order in $r$.

$$e^{-\sigma^2 r^2} \approx 1 - \sigma^2 r^2, \qquad \sigma r \ll 1 \tag{15}$$

Inserting Eq.(15) into (14) we obtained

$$\frac{d^2\psi}{dr^2} + \left(\epsilon_{nl} + \frac{a}{r} - cr + dr^2 - \frac{L}{r^2}\right)\psi(r) = 0 \tag{16}$$

where

$$a = \frac{8\mu\alpha_s}{3\hbar^2}, c = \frac{2\mu b}{\hbar^2}, d = \frac{32\mu\sigma^2 \alpha_s \pi\left(\frac{\sigma}{\sqrt{\pi}}\right)^3\left(s(s+1)-\frac{3}{2}\right)}{9\hbar^2 m_q m_{\bar{q}}}$$

$$\epsilon_{nl} = \frac{2\mu E_{nl}}{\hbar^2} - \frac{d}{\sigma^2}, \quad L = l(l+1)$$

Using coordinate transformation, $= \frac{1}{q}$ , Eq. (16) reduces to

$$\frac{d^2\psi}{dq^2} + \frac{2q}{q^2}\frac{d\psi}{dq} + \frac{1}{q^4}\left(\epsilon_{nl} + aq - \frac{c}{q} + \frac{d}{q^2} - Lq^2\right)\psi(q) = 0 \tag{17}$$

It is possible to solve Eq. (17) using the fractional Nikiforov-Uvarov method shown in Section (2). However, we must expand $\frac{c}{q}$ and $\frac{d}{q^2}$ in a power series to second-order around $r_0$ $\left(\delta = \frac{1}{r_0}\right)$ in q-space which is assume to be the characteristics radius of the mesons and use the approximation method to convert the equation to a solvable form. This is comparable to the Pekeris approximation [55], which aids in deforming the centrifugal potential such that the changed potential may be calculated using the NU technique.

Putting $y = q - \delta$ and around y = 0, so that we can write

$$\frac{c}{q} = \frac{c}{(y+\delta)} \approx \frac{c}{\delta}\left(1 - \frac{y}{\delta} + \frac{y^2}{\delta^2}\right) = c\left(\frac{3}{\delta} - \frac{3q}{\delta^2} + \frac{q^2}{\delta^3}\right) \tag{18}$$

$$\frac{d}{q^2} = \frac{d}{(y+\delta)^2} \approx \frac{d}{\delta^2}\left(1 - \frac{2y}{\delta} + \frac{3y^2}{\delta^2}\right) = d\left(\frac{6}{\delta^2} - \frac{8q}{\delta^3} + \frac{3q^2}{\delta^4}\right) \tag{19}$$

We note that Eqs. (18, 19) give a good accuracy when $y$ tends to zero means q $\approx$ $\delta$ as seen Fig. 2.

Inserting (18) and (19) into (17) results to

$$\frac{d^2\psi}{dq^2} + \frac{2q}{q^2}\frac{d\psi}{dq} + \frac{1}{q^4}(-Aq^2 + Bq - C)\psi(q) = 0 \tag{20}$$

where

$$A = \frac{c}{\delta^3} - \frac{3d}{\delta^4} + L \tag{21}$$

$$B = a + \frac{3c}{\delta^2} - \frac{8d}{\delta^3} \tag{22}$$

$$C = -\epsilon_{nl} + \frac{3c}{\delta} - \frac{6d}{\delta^2} \tag{23}$$

Eq. (20) is written in the fractional form of dimensionless units by taking $q = s\,\mu_1$ where s is the dimensionless variable and $\mu$ is scale parameter equals 1 GeV, so we can write Eq. (20)

$$D^\alpha D^\alpha \psi(s) + \frac{\tilde{\tau}(s)}{\sigma(s)} D^\alpha \psi(s) + \frac{\tilde{\sigma}(s)}{\sigma^2(s)} \psi(s) = 0 \tag{24}$$

where

$$\tilde{\tau}(s) = 2s^\alpha, \sigma(s) = s^{2\alpha} \text{ and } \tilde{\sigma}(s) = -As^{2\alpha} + \frac{B}{\mu_1}s^\alpha - \frac{C}{\mu_1} \tag{25}$$

From eq. (5), we have $\pi(qs)$ as

$$\pi(q) = \frac{(2\alpha k_1 - 2)s^\alpha}{2} \pm \sqrt{\left(\frac{(2\alpha k_1 - 2)^2}{4} + A + k\right)s^{2\alpha} - \frac{B}{\mu_1}s^\alpha + \frac{C}{\mu_1}} \tag{26}$$

We note that $\pi(qs)$ is a first-order polynomial, consequently the square root's terms must be equal to the square of a linear function, say, $(Xq + Y)^2$. With this, we have

$(Xs^\alpha + Y)^2 = \left(\frac{(2\alpha k_1 - 2)^2}{4} + A + k\right)s^{2\alpha} - \frac{B}{\mu_1}s^\alpha + \frac{C}{\mu_1}$ and solving it gives

$$k = \frac{B^2}{4\mu C} - \frac{(2\alpha k_1 - 2)^2}{4} - A \tag{27}$$

Inserting (27) into (26) gives

$$\pi(s) = \frac{(2\alpha k_1 - 2)^2}{2}s^\alpha \pm \left[\pm\left(\frac{Bs^\alpha}{2\sqrt{\mu C}} - \sqrt{\frac{C}{\mu}}\right)\right] \tag{28}$$

In (28), $\pi(s) = \frac{(2\alpha k_1 - 2)^2}{2} s^\alpha - \left[\left(\frac{Bs^\alpha}{2\sqrt{\mu C}} - \sqrt{\frac{C}{\mu_1}}\right)\right]$ that will make $\frac{d\tau(q)}{dq} < 0$

The energy eigenvalue in (8) is

$$\sqrt{\mu_1 C} = \frac{B}{2\left(n\alpha k_1 + \frac{\alpha k_1}{2} \pm \sqrt{\left(n\alpha k_1 + \frac{\alpha k_1}{2}\right)^2 - F}\right)} \tag{29}$$

where

$$F = n(n-1)k_1^2\alpha^2 + 2nk_1\alpha + nk_1\alpha(2k_1\alpha - 2) - \frac{(2k_1\alpha - 2)^2}{4} - A + \frac{k_1\alpha(2k_1\alpha - 2)}{2} \tag{30}$$

Injecting the constants $A, B$ and $C$ in (21)-(23) into Eq.(29) yields

$$\epsilon_{nl} = \Delta_0 - \frac{1}{4}\left(\frac{\Delta_1}{(n\alpha k_1 + \Delta_2)}\right)^2 \tag{31}$$

where

$$\Delta_0 = \frac{3c}{\delta} - \frac{6d}{\delta^2} \tag{32}$$

$$\Delta_1 = a + \frac{3c}{\delta^2} - \frac{8d}{\delta^3} \tag{33}$$

$$\Delta_2 = n\alpha k_1 + \frac{\alpha k_1}{2} \pm \sqrt{\left(n\alpha k_1 + \frac{\alpha k_1}{2}\right)^2 - F} \tag{34}$$

Using the parameters for $a, c, d$ and $L$ in (16), the ES of the HM system is obtained as

$$E_{nl} = \Gamma_0 - \frac{\mu}{2\hbar^2}\left(\frac{\Gamma_1}{n\alpha k_1 + \Gamma_2}\right)^2 \tag{35}$$

where

$$\Gamma_0 = \frac{3b}{\delta} + \frac{16\alpha_s \pi \left(\frac{\sigma}{\sqrt{\pi}}\right)^3 \left(s(s+1) - \frac{3}{2}\right)}{9 m_q m_{\bar{q}}} \left[1 - \frac{6\sigma^2}{\delta^2}\right]$$

$$\Gamma_1 = \frac{4\alpha_s}{3} + \frac{3b}{\delta^2} - \frac{128\alpha_s \sigma^2 \pi \left(\frac{\sigma}{\sqrt{\pi}}\right)^3 \left(s(s+1) - \frac{3}{2}\right)}{9 m_q m_{\bar{q}} \delta^3}$$

$$\Gamma_2 = \frac{\alpha k_1}{2} \pm \sqrt{\left(n\alpha k_1 + \frac{\alpha k_1}{2}\right)^2 - \Gamma_3}$$

where

$$\Gamma_3 = n(n-1)k_1^2\alpha^2 + 2nk_1\alpha + nk_1\alpha(2k_1\alpha - 2) - \frac{(2k_1\alpha - 2)^2}{4} - \frac{2\mu b}{\hbar^2 \delta^3}$$

$$+ \frac{96\alpha_s \sigma^2 \pi \left(\frac{\sigma}{\sqrt{\pi}}\right)^3 \left(s(s+1) - \frac{3}{2}\right)}{9 m_q m_{\bar{q}} \delta^4} - \left(l + \frac{1}{2}\right)^2 + \frac{1}{4}$$

The wave function (WF) is determined from weight functions as

$$\rho(s) = s^{s_1} e^{-\frac{2\sqrt{\frac{C}{\mu_1}}}{\alpha} s^{-\alpha}} \tag{36}$$

where

$$s_1 = -2\alpha + \frac{2}{k_1} + \frac{(2\alpha k_1 - 2)}{2} - \frac{B}{k_1\sqrt{C}}$$

$$\chi_n(s) = B_{nl} s^{-s_1} e^{\frac{2\sqrt{\frac{C}{\mu_1}}}{\alpha} s^{-\alpha}} (D^\alpha)^n \left( s^{2n\alpha + s_1} e^{-\frac{2\sqrt{\frac{C}{\mu_1}}}{\alpha} s^{-\alpha}} \right) \tag{37}$$

The other part of the WF is obtained using Eq. (12)

$$\phi(s) = s^{s_2} e^{-\frac{\sqrt{\frac{C}{\mu}}}{\alpha k_1} s^{-\alpha}} \tag{38}$$

where

$$s_2 = \frac{(2\alpha k_1 - 2)}{2k_1} - \frac{B}{2k_1\sqrt{\mu C}}$$

Using (37) and (38) the total WF is gotten as

$$\psi_n(s) = B_{nl} s^{-s_1 + s_2} e^{\frac{2\sqrt{\frac{C}{\mu_1}}}{\alpha} s^{-\alpha} - \frac{\sqrt{\frac{C}{\mu}}}{\alpha k_1} s^{-\alpha}} (D^\alpha)^n \left( s^{2n\alpha + s_1} e^{-\frac{2\sqrt{\frac{C}{\mu_1}}}{\alpha} s^{-\alpha}} \right) \qquad s = \frac{1}{\mu_1 r} \tag{39}$$

The normalization constant can be obtain using the relation

$$\int_0^\infty |\psi_{nl}(r)|^2 dr = 1 \tag{40}$$

## 4. Discussion of numerical results

In section (3), we have obtained the energy eigenvalue and wave function in the fractional forms as in Eqs. (35) and (39). We have obtained the results of Ref. [52] at $\alpha = \beta = 1 \rightarrow k_1 = 1$ $and$ $\mu_1 = 1$

$$E_{nl} = \Gamma_0 - \frac{\mu}{2\hbar^2} \left( \frac{\Gamma_1}{n + \Gamma_2} \right)^2 \tag{41}$$

where

$$\Gamma_0 = \frac{3b}{\delta} + \frac{16\alpha_s \pi \left(\frac{\sigma}{\sqrt{\pi}}\right)^3 \left(s(s+1) - \frac{3}{2}\right)}{9 m_q m_{\bar{q}}} \left[1 - \frac{6\sigma^2}{\delta^2}\right]$$

$$\Gamma_1 = \frac{4\alpha_a}{3} + \frac{3b}{\delta^2} - \frac{128 \alpha_s \sigma^2 \pi \left(\frac{\sigma}{\sqrt{\pi}}\right)^3 \left(s(s+1) - \frac{3}{2}\right)}{9 m_q m_{\bar{q}} \delta^3}$$

$$\Gamma_2 = \frac{1}{2} \pm \sqrt{\frac{2\mu b}{\hbar^2 \delta^3} - \frac{96\alpha_s \sigma^2 \pi \left(\frac{\sigma}{\sqrt{\pi}}\right)^3 \left(s(s+1)-\frac{3}{2}\right)}{9 m_q m_{\bar{q}} \delta^4} + \left(l + \frac{1}{2}\right)^2}$$

The wave function is given by

$$\psi_n(q) = N_{nl} q^{\frac{B}{2\sqrt{C}}} e^{\frac{\sqrt{C}}{q}} \frac{d^n}{dq^n}\left(q^{2n-\frac{B}{\sqrt{C}}} e^{-2\frac{\sqrt{C}}{q}}\right) \qquad q = \frac{1}{r} \qquad (42)$$

To calculate the mass spectra for the heavy mesons, we used the energy mass relations

$$M_{nl} = m_q + m_{\bar{q}} + E_{nl} \qquad (43)$$

The potential free parameters for arbitrary fractional values were obtained by fitting the analytical equation with the experimental data of the particle Data Group [53]. We used fixed values of bottom and charm quark masses which had been measured as $1.2 < m_c < 1.8 GeV$ and $4.5 < m_b < 5.4 GeV$ respectively [25]. The free parameters of the spin-dependent potential with fixed values of fractional terms and masses are given in Table 1. For the bottom-charm meson due to lack of available experimental data, we have assumed a constant characteristic radius ($\delta$) and smearing constant $\sigma$ to obtain other unknown constants
$\alpha_s$ and $b$. In addition, the parameters were used to compute the masses of the heavy mesons given in Table 2, Table 3 and Table 4. The masses for the triplet (s=1) and singlet states (s=0) have been obtained for different quantum S, P, D, F and G states.

**Table 1. Calculated potential parameters for the mesons.**

| Parameters | Bottomonium $m_b = m_{\bar{b}} = 4.890$ | | Charmonium $m_c = m_{\bar{c}} = 1.27$ | | Bottom-charm | |
|---|---|---|---|---|---|---|
| Mass(GeV) | $\alpha = 0.97, \beta = 1$ | $\alpha = 0.5, \beta = 1$ | $\alpha = 0.97, \beta = 1$ | $\alpha = 0.5, \beta = 1$ | $\alpha = 0.97, \beta = 1$ | $\alpha = 0.5, \beta = 1$ |
| $\alpha_s$ | 1.6854 | 1.2918 | 2.8391 | 1.8164 | 2.1208 | 1.8289 |
| $\delta(GeV)$ | 0.7610 | 0.7153 | 0.5910 | 0.4882 | 0.5910 | 0.4882 |
| $b(GeV^2)$ | 0.3200 | 0.3884 | 0.4281 | 0.4175 | 0.2931 | 0.3711 |
| $\sigma(GeV)$ | 0.5384 | 0.6586 | 0.3021 | -0.2969 | 0.3021 | -0.2969 |

In Table 2, the bottomonium mass spectra with different values of fractional parameters $\beta = 1, \alpha = 0.97$ and $\beta = 1, \alpha = 0.50$ have been obtained for the s-wave and $l > 0$ quantum states. Generally, the bottomonium masses increase with the increase in the quantum number from $n = 1 - 6$ for both the singlet ($s = 0$) and triplet ($s = 1$) quantum states. Also, a similar trend is observed for the and $l > 0$ quantum states. The bottomonium masses are in good agreement with the masses obtained in existing literature [27, 28, 33-36] and available experimental data [53]. The percentage errors deviation from experimental data for $\beta = 1, \alpha = 0.97$ and $\beta = 1, \alpha = 0.50$ were found to be 0.67% and 0.49% respectively. The deviations give improvement over the previous results obtained in [52] where the authors solved the SE without considering the fractional parameters. Also, the results for bottomonium mass spectra give improvement over the results reported in Ref. [35]. It is observed that the smaller the value of $\alpha$ in the fractional Schrodinger equation the better the results. This is evident when $\alpha = 0.50$ yielding a 0.49% error which is comparable to the percentage errors of 0.33% [34] and 0.40% [37] obtained in previous works. Furthermore, Using potential parameters for charmonium meson given in Table 1, we obtained the mass spectra for different quantum states as tabulated in Table 3. The charmonium masses increase as the quantum number increases for spin-singlet ($s = 0$) and triplet (s=1) states. In comparison to experimental data [53], a percentage error 1.97% and 1.62% were obtained for fractional values of $\beta = 1, \alpha = 0.97$ and $\beta = 1, \alpha = 0.50$ respectively. For small fractional parameters, the results give improvements over previous studies [52] and literature data [35]. The results further indicated that the potential parameters coupled with the adjustment of the fractional parameters in the Schrödinger equation strongly correlate with the accuracy of the heavy meson mass spectra. In Table 4, we obtained the mass spectra of the bottom-charm meson for the s-wave and $l > 0$ states for the spin number $s = 0$ and $s = 1$. The masses also follow the trends of the bottomonium and charmonium and conform to the results reported in existing literature [27, 35, 36]. In Figure 1, (a-c), we plotted the potential against radial distance using the calculated parameters for the bottomonium and charmonium mesons in Table 1. It can be seen that the potential curves account for the attractive Coulomb and linear confinement which have been used to describe the non-relativistic quarkonium interaction within the quantum chromo-dynamic and phenomenology approach. In Figure 2 (a-d), we graphed the Pekeris-type approximation given in equations (18) and (19) in the region $q = \delta$ which gives a fair approximation compared to the exact curve.

**Table 2**. Bottomonium mass spectra in $GeV$ under spin-spin interaction potential.

| State | | [52] | Present | | [34] | [28] | [36] | [27] | [33] | [35] | Expt.[53] |
|---|---|---|---|---|---|---|---|---|---|---|---|
| $n^{2s+1}L_j$ | $J^{PC}$ | | $\alpha = 0.97$ | $\alpha = 0.5$ | | | | | | | |
| $\Upsilon(1^3S_1)^*$ | $1^{--}$ | 9.460 | 9.460 | 9.497 | 9.463 | 9.465 | 9.460 | 9.525 | 9.460 | 9.460 | $9.460 \pm (2.6 \times 10^{-4})$ |
| $\eta_b(1^1S_0)^*$ | $0^{-+}$ | 9.399 | 9.437 | 9.419 | 9.423 | 9.402 | 9.398 | 9.472 | 9.390 | 9.428 | $9.399 \pm (2 \times 10^{-3})$ |
| $\Upsilon(2^3S_1)$ | $1^{--}$ | 10.064 | 10.084 | 10.023 | 10.001 | 10.003 | 10.023 | 10.049 | 10.015 | 9.979 | $10.023 \pm (3.1 \times 10^{-4})$ |
| $\eta_b(2^1S_0)$ | $0^{-+}$ | 10.074 | 10.084 | 9.993 | 9.983 | 9.976 | 9.990 | 10.028 | 9.990 | 9.955 | |
| $\Upsilon(3^3S_1)^*$ | $1^{--}$ | 10.355 | 10.399 | 10.355 | 10.354 | 10.354 | 10.355 | 10.371 | 10.343 | 10.359 | $10.355 \pm (5 \times 10^{-4})$ |
| $\eta_b(3^1S_0)$ | $0^{-+}$ | 10.408 | 10.412 | 10.357 | 10.342 | 10.336 | 10.329 | 10.360 | 10.326 | 10.338 | |
| $\Upsilon(4^3S_1)$ | $1^{--}$ | 10.517 | 10.579 | 10.579 | 10.650 | 10.635 | 10.586 | 10.598 | 10.597 | 10.683 | $10.579 \pm (1.2 \times 10^{-3})$ |
| $\eta_b(4^1S_0)$ | $0^{-+}$ | 10.596 | 10.600 | 10.606 | 10.638 | 10.623 | 10.573 | 10.592 | 10.584 | 10.663 | |
| $\Upsilon(5^3S_1)$ | $1^{--}$ | 10.617 | 10.692 | 10.736 | 10.912 | 10.878 | 10.851 | 10.870 | 10.811 | 10.975 | $10.885^{+2.6 \times 10^{-3}}_{-1.6 \times 10^{-3}}$ |
| $\eta_b(5^1S_0)$ | $0^{-+}$ | 10.713 | 10.718 | 10.781 | 10.901 | 10.869 | 10.869 | 10.790 | 10.800 | 10.956 | |
| $\Upsilon(6^3S_1)$ | $1^{--}$ | 10.682 | 10.767 | 10.852 | 11.150 | 11.102 | 11.061 | 11.022 | 10.997 | 11.243 | $11.020 \pm (4 \times 10^{-3})$ |
| $\eta_b(6^1S_0)$ | $0^{-+}$ | 10.791 | 10.797 | 10.910 | 11.140 | 11.097 | 11.088 | 10.961 | 10.988 | 11.226 | 11.014 |
| $\chi_{b_1}(1^3P_1)$ | $1^{++}$ | 9.827 | 9.810 | 9.893 | 9.894 | 9.876 | 9.892 | 9.875 | 9.903 | 9.819 | $9.893 \pm (2.6 \times 10^{-4}) \pm (3.1 \times 10^{-4})$ |
| $h_b(1^1P_1)$ | $1^{+-}$ | 9.789 | 9.796 | 9.842 | 9.899 | 9.882 | 9.900 | 9.884 | 9.909 | 9.821 | $9.899 \pm (8 \times 10^{-4})$ |
| $\chi_{b_1}(2^3P_1)$ | $1^{++}$ | 10.235 | 10.255 | 10.271 | 10.265 | 10.246 | 10.255 | 10.229 | 10249 | 10.2170 | $10.255 \pm (2.2 \times 10^{-4}) \pm (5 \times 10^{-4})$ |
| $h_b(2^1P_1)^*$ | $1^{+-}$ | 10.260 | 10.260 | 10.260 | 10.268 | 10.250 | 10.260 | 10.237 | 10254 | 10.220 | $10.260 \pm (1.2 \times 10^{-3})$ |
| $\chi_{b_1}(3^3P_1)$ | $1^{++}$ | 10.448 | 10.494 | 10.521 | 10.567 | 10.538 | 10.541 | 10.339 | 10.515 | 10.553 | $10.5135 \pm (7 \times 10^{-4})$ |
| $h_b(3^1P_1)$ | $1^{+-}$ | 10.510 | 10.510 | 10.538 | 10.570 | 10.541 | 10.544 | 10.362 | 10.519 | 10.556 | |
| $\chi_{b_1}(4^3P_1)$ | $1^{++}$ | 10.573 | 10.637 | 10.695 | | 10.788 | 10.802 | 10.571 | | | |
| $h_b(4^1P_1)$ | $1^{+-}$ | 10.659 | 10.661 | 10.732 | | 10.790 | 10.804 | 10.594 | | | , |
| $\Upsilon_2(1^3D_2)^*$ | $2^{--}$ | 10.164 | 10.164 | 10.319 | 10.149 | 10.147 | 10.161 | 10.096 | 10.153 | 10.075 | $10.164 \pm (1.4 \times 10^{-3})$ |
| $\eta_{b_2}(1^1D_2)$ | $2^{-+}$ | 10.167 | 10.163 | 10.307 | 10.150 | 10.148 | 10.163 | 9.767 | 10.153 | 10.074 | |
| $\Upsilon_2(2^3D_2)$ | $2^{--}$ | 10.295 | 10.443 | 10.554 | 10.465 | 10.449 | 10.443 | 10.071 | 10.432 | 10.424 | |
| $\eta_{b_2}(2^1D_2)$ | $2^{-+}$ | 10.407 | 10.455 | 10.570 | 10.465 | 10.450 | 10.445 | 10.093 | 10.432 | 10.424 | |
| $\Upsilon_2(3^3D_2)$ | $2^{--}$ | 10.549 | 10.606 | 10.718 | 10.740 | 10.705 | 10.711 | 10.345 | | 10.733 | |
| $\eta_{b_2}(3^1D_2)$ | $2^{-+}$ | 10.627 | 10.627 | 10.756 | 10.740 | 10.706 | 10.713 | 10.368 | | 10.733 | |
| $\chi_{b_3}(1^3F_3)$ | $3^{++}$ | 10.386 | 10.412 | 10.628 | | 10.355 | 10.346 | 9.754 | 10.340 | 10.287 | |
| $\eta_{b_3}(1^1F_3)$ | $3^{+-}$ | 10.427 | 10422 | 10.651 | | 10.355 | 10.347 | 9.778 | 10.339 | 10.288 | |
| $\chi_{b_3}(2^3F_3)$ | $3^{++}$ | 10.535 | 10.587 | 10.772 | | 10.619 | 10.614 | 10.081 | | 10.607 | |
| $\eta_{b_3}(2^1F_3)$ | $3^{+-}$ | 10.608 | 10.606 | 10.814 | | 10.619 | 10.615 | 10.104 | | 10.607 | |
| $\chi_{b_4}(1^3G_4)$ | $4^{--}$ | 10.527 | 10.576 | 10.833 | | 10.531 | 10.512 | | | | |
| $\eta_{b_4}(1^1G_4)$ | $4^{-+}$ | 10.595 | 10.594 | 10.882 | | 10.530 | 10.513 | | | | |
| % Error | | 075 | 0.67 | 0.49 | 0.33 | 0.24 | 0.13 | 0.40 | 0.15 | 0.72 | |

**Table 3.** Charmonium mass spectra in $GeV$ under spin-spin interaction potential.

| State | | Present | | [52] | [35] | [32] LP(SP) | [23] | [36] | [27] | Expt. [53] |
|---|---|---|---|---|---|---|---|---|---|---|
| $n^{2s+1}L_j$ | $J^{PC}$ | $\alpha=0.97$ | $\alpha=0.5$ | | | | | | | |
| $J/\psi(1^3S_1)^*$ | $1^{--}$ | 2.950 | 3.097 | 3.081(3.097) | 3.094 | 3.097(3.097) | 3.094 | 3.096 | 3.126 | $3.097 \pm (6 \times 10^{-6})$ |
| $\eta_c(1^1S_0)^*$ | $0^{-+}$ | 2.984 | 3.045 | 3.061(2.979) | 2.989 | 2.983(2.984) | 2.995 | 2.981 | 3.033 | $2.984 \pm 0.0004$ |
| $\psi(2^3S_1)$ | $1^{--}$ | 3.686 | 3.677 | 3.717(3.736) | 3.681 | 3.679(3.679) | 3.649 | 3.685 | 3.701 | $3.686 \pm (6 \times 10^{-5})$ |
| $\eta_c(2^1S_0)$ | $0^{-+}$ | 3.709 | 3.638 | 3.696(3.640) | 3.602 | 3.635(3.637) | 3.606 | 3.635 | 3.666 | $3.638 \pm (1.1 \times 10^{-3})$ |
| $\psi(3^3S_1)^*$ | $1^{--}$ | 4.039 | 4.039 | 4.003(4.039) | 4.129 | 4.078(4.030) | 4.036 | 4.039 | 4.055 | $4.039 \pm 10^{-3}$ |
| $\eta_c(3^1S_0)$ | $0^{-+}$ | 4.066 | 4.002 | 3.983(3.937) | 4.058 | 4.048(4.004) | 4.000 | 3.989 | 4.158 | |
| $\psi(4^3S_1)$ | $1^{--}$ | 4.235 | 4.280 | 4.156(4.206) | 4.514 | 4.412(4.281) | 4.362 | 4.427 | 4.415 | $4.421 \pm (4 \times 10^{-3})$ |
| $\eta_c(4^1S_0)$ | $0^{-+}$ | 4.267 | 4.242 | 4.137(4.096) | 4.448 | 4.388(4.264) | 4.328 | 4.401 | 4.415 | |
| $\psi(5^3S_1)$ | $1^{--}$ | 4.355 | 4.449 | 4.247(4.308) | 4.863 | 4.711(4.472) | 4.654 | 4.837 | 4.585 | $4.63 \pm (6 \times 10^{-3})$ |
| $\eta_c(5^1S_0)$ | $0^{-+}$ | 4.391 | 4.408 | 4.228(4.190) | 4.799 | 4.690(4.459) | 4.622 | 4.811 | 4.607 | |
| $\psi(6^3S_1)$ | $1^{--}$ | 4.434 | 4.572 | 4.305(4.374) | 5.185 | | 4.925 | 5.167 | 4.733 | |
| $\eta_c(6^1S_0)$ | $0^{-+}$ | 4.473 | 4.527 | 4.287(4.250) | 5.124 | | 4.893 | 5.155 | 4.754 | |
| $\chi_{c_1}(1^3P_1)$ | $1^{++}$ | 3.412 | 3.547 | 3.528(3.501) | 3.468 | 3.516(3.521) | 3.523 | 3.511 | 3.487 | $3.511 \pm (5 \times 10^{-5})$ |
| $h_c(1^1P_1)^*$ | $1^{+-}$ | 3.416 | 3.525 | 3.503(3.446) | 3.470 | 3.522(3.526) | 3.534 | 3.525 | 3.502 | $3.525 \pm (1.1 \times 10^{-4})$ |
| $\chi_{c_1}(2^3P_1)^*$ | $1^{++}$ | 3.900 | 3.955 | 3.912(3.921) | 3.938 | 3.937(3.914) | 3.925 | 3.906 | 3.786 | $3.872 \pm (6 \times 10^{-5})$ |
| $h_c(2^1P_1)$ | $1^{+-}$ | 3.915 | 3.931 | 3.889(3.843) | 3.943 | 3.940(3.916) | 3.936 | 3.926 | 3.821 | |
| $\chi_{c_1}(3^3P_1)$ | $1^{++}$ | 4.155 | 4.224 | 4.105(4.139) | 4.338 | 4.284(4.192) | 4.257 | 4.319 | 4.123 | $4.147 \pm (3 \times 10^{-3})$ |
| $h_c(3^1P_1)$ | $1^{+-}$ | 4.179 | 4.194 | 4.084(4.043) | 4.344 | 4.285(4.193) | 4.269 | 4.337 | 4.164 | |
| $\chi_{c_1}(4^3P_1)$ | $1^{++}$ | 4.305 | 4.409 | 4.215(4.266) | 4.696 | | | 4.728 | 4.373 | |
| $h_c(4^1P_1)$ | $1^{+-}$ | 4.335 | 4.374 | 4.196(4.158) | 4.704 | | | 4.744 | 4.420 | |
| $\psi_2(1^3D_2)$ | $2^{--}$ | 3.823 | 4.015 | 3.868(3.855) | 3.772 | 3.807(3.807) | 3.805 | 3.795 | 3.348 | $3.823 \pm (5 \times 10^{-4})$ |
| $\eta_{c_2}(1^1D_2)$ | $2^{-+}$ | 3.825 | 4.001 | 3.844(3.798) | 3.765 | 3.806(3.805) | 3.802 | 3.807 | 3.376 | |
| $\psi_2(2^3D_2)$ | $2^{--}$ | 4.113 | 4.264 | 4.081(4.102) | 4.188 | 4.165(4.109) | 4.152 | 4.190 | 3.801 | |
| $\psi_2(2^1D_2)$ | $2^{-+}$ | 4.129 | 4.241 | 4.059(4.019) | 4.182 | 4.164(4.108) | 4.150 | 4.196 | 3.836 | |
| $\psi_2(3^3D_2)$ | $2^{--}$ | 4.279 | 4.438 | 4.201(4.243) | 4.557 | 4.478(4.337) | 4.456 | 4.544 | 4.135 | |
| $\eta_{c_2}(3^1D_2)$ | $2^{-+}$ | 4.304 | 4.407 | 4.181(4.143) | 4.553 | 4.478(4.336) | 4.455 | 4.549 | 4.176 | |
| $\chi_{c_3}(1^3F_3)$ | $3^{++}$ | 4.089 | 4.346 | 4.068(4.081) | 4.012 | | | 4.068 | 3.375 | |
| $\eta_{c_3}(1^1F_3)$ | $3^{+-}$ | 4.100 | 4.325 | 4.046(4.006) | 4.017 | | | 4.071 | 3.403 | |
| $\chi_{c_3}(2^3F_3)$ | $3^{++}$ | 4.265 | 4.497 | 4.194(4.231) | 4.396 | | | 4.400 | 3.823 | |
| $\eta_{c_3}(2^1F_3)$ | $3^{+-}$ | 4.287 | 4.467 | 4.173(4.135) | 4.400 | | | 4.406 | 3.858 | |
| $\chi_{c_4}(1^3G_4)$ | $4^{--}$ | 4.257 | 4.562 | 4.189(4.223) | | | | 4.343 | | |
| $\eta_{c_4}(1^1G_4)$ | $4^{-+}$ | 4.277 | 4.531 | 4.168(4.131) | | | | 4.345 | | |
| % Error | | 1.97% | 1.62% | 2.09(1.52) | 1.77 | 0.74(0.83) | 0.81 | 0.90 | 0.85 | |

**Table 4.** Bottom-charm mass spectra in $GeV$ under spin-spin interaction potential.

| State | | Present | | [52] | [36] | [1] | [9] | Expt.[54] |
|---|---|---|---|---|---|---|---|---|
| $n^{2s+1}L_j$ | $J^{PC}$ | $\alpha=0.97$ | $\alpha=0.5$ | | | | | |
| $1^3S_1$ | $1^{--}$ | 6.278 | 6.272 | 6.2808 | 6.3330 | 6.313 | 6.320 | |
| $(1^1S_0)^*$ | $0^{-+}$ | 6.275 | 6.275 | 6.2750 | 6.2720 | 6.276 | 6.721 | 6.275 |
| $2^3S_1$ | $1^{--}$ | 6.842 | 6.841 | 6.8523 | 6.8820 | 6.867 | 6.900 | |
| $(2^1S_0)^*$ | $0^{-+}$ | 6.842 | 6.842 | 6.8420 | 6.8420 | 6.841 | 6.864 | 6.842 |
| $3^3S_1$ | $1^{--}$ | 7.118 | 7.213 | 7.1179 | 7.2580 | 7.308 | 7.338 | |
| $3^1S_0$ | $0^{-+}$ | 7.120 | 7.211 | 7.1027 | 7.2260 | 7.281 | 7.306 | |
| $4^3S_1$ | $1^{--}$ | 7.272 | 7.469 | 7.2626 | 7.6090 | 7.660 | 7.714 | |
| $4^1S_0$ | $0^{-+}$ | 7.276 | 7.465 | 7.2437 | 7.5850 | 7.634 | 7.684 | |
| $5^3S_1$ | $1^{--}$ | 7.368 | 7.653 | 7.3500 | 7.9470 | 7.941 | 8.054 | |
| $5^1S_0$ | $0^{-+}$ | 7.373 | 7.647 | 7.3284 | 7.9280 | 7.917 | 8.025 | |
| $6^3S_1$ | $1^{--}$ | 7.431 | 7.789 | 7.4068 | | 8.168 | 8.368 | |
| $6^1S_0$ | $0^{-+}$ | 7.437 | 7.782 | 7.3833 | | 8.144 | 8.340 | |
| $1^3P_1$ | $1^{++}$ | 6.619 | 6.664 | 6.6570 | 6.743 | 6.281 | 6.705 | |
| $1^1P_1$ | $1^{+-}$ | 6.614 | 6.669 | 6.6567 | 6.750 | 6.290 | 6.706 | |
| $2^3P_1$ | $1^{++}$ | 7.003 | 7.095 | 7.0213 | 7.134 | 6.836 | 7.165 | |
| $2^1P_1$ | $1^{+-}$ | 7.003 | 7.096 | 7.0117 | 7.147 | 6.846 | 7.168 | |
| $3^3P_1$ | $1^{++}$ | 7.206 | 7.386 | 7.2079 | 7.500 | 7.278 | 7.555 | |
| $3^1P_1$ | $1^{+-}$ | 7.208 | 7.385 | 7.1924 | 7.510 | 7.287 | 7.559 | |
| $4^3P_1$ | $1^{++}$ | 7.326 | 7.593 | 7.3161 | 7.844 | 7.631 | 7.905 | |
| $4^1P_1$ | $1^{+-}$ | 7.330 | 7.589 | 7.2968 | 7.853 | 7.640 | 7.908 | |
| $1^3D_2$ | $2^{--}$ | 6.936 | 7.125 | 6.9697 | 7.025 | 6.299 | 6.997 | |
| $1^1D_2$ | $2^{-+}$ | 6.934 | 7.128 | 6.9646 | 7.026 | 6.308 | 6.994 | |
| $2^3D_2$ | $2^{--}$ | 7.169 | 7.407 | 7.1797 | 7.399 | 6.852 | 7.403 | |
| $2^1D_2$ | $2^{-+}$ | 7.170 | 7.407 | 7.1668 | 7.400 | 6.861 | 7.401 | |
| $3^3D_2$ | $2^{--}$ | 7.303 | 7.608 | 7.2991 | 7.741 | 7.290 | 7.764 | |
| $3^1D_2$ | $2^{-+}$ | 7.306 | 7.605 | 7.2813 | 7.743 | 7.302 | 7.762 | |
| $1^3F_3$ | $3^{++}$ | 7.148 | 7.484 | 7.1639 | 7.269 | 6.326 | 7.242 | |
| $1^1F_3$ | $3^{+-}$ | 7.148 | 7.484 | 7.1527 | 7.268 | 6.335 | 7.241 | |
| $2^3F_3$ | $3^{++}$ | 7.290 | 7.664 | 7.2896 | 7.616 | 6.876 | 7.615 | |
| $2^1F_3$ | $3^{+-}$ | 7.293 | 7.661 | 7.2729 | 7.615 | 6.885 | 7.614 | |
| $1^3G_4$ | $4^{--}$ | 7.283 | 7.734 | 7.3635 | 7.489 | | | |
| $1^1G_4$ | $4^{-+}$ | 7.285 | 7.731 | 7.3437 | 7.487 | | | |

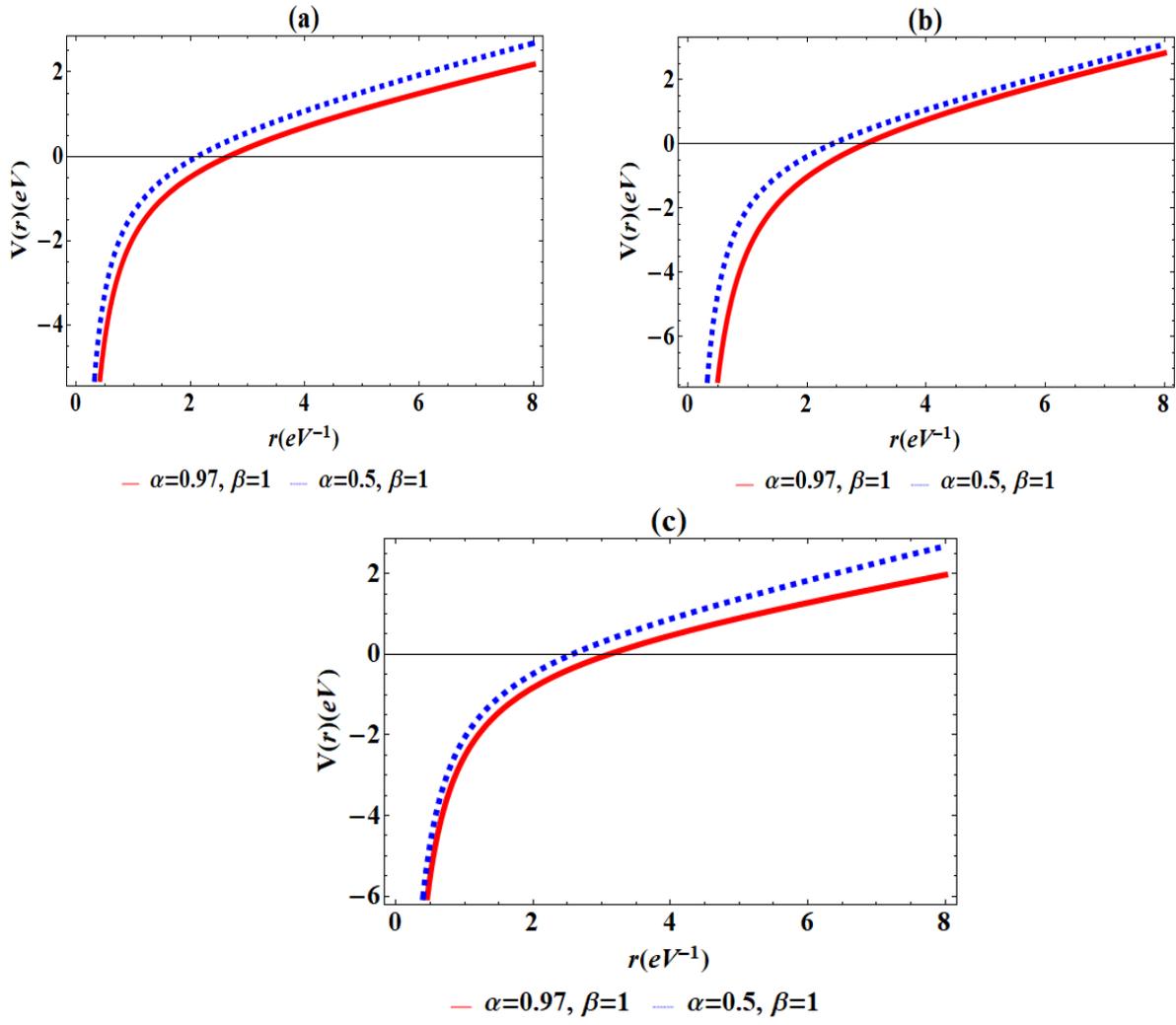

Figure 1. Potential curves for (a) bottomonium. (b) charmonium. (c) bottom-charm mesons.

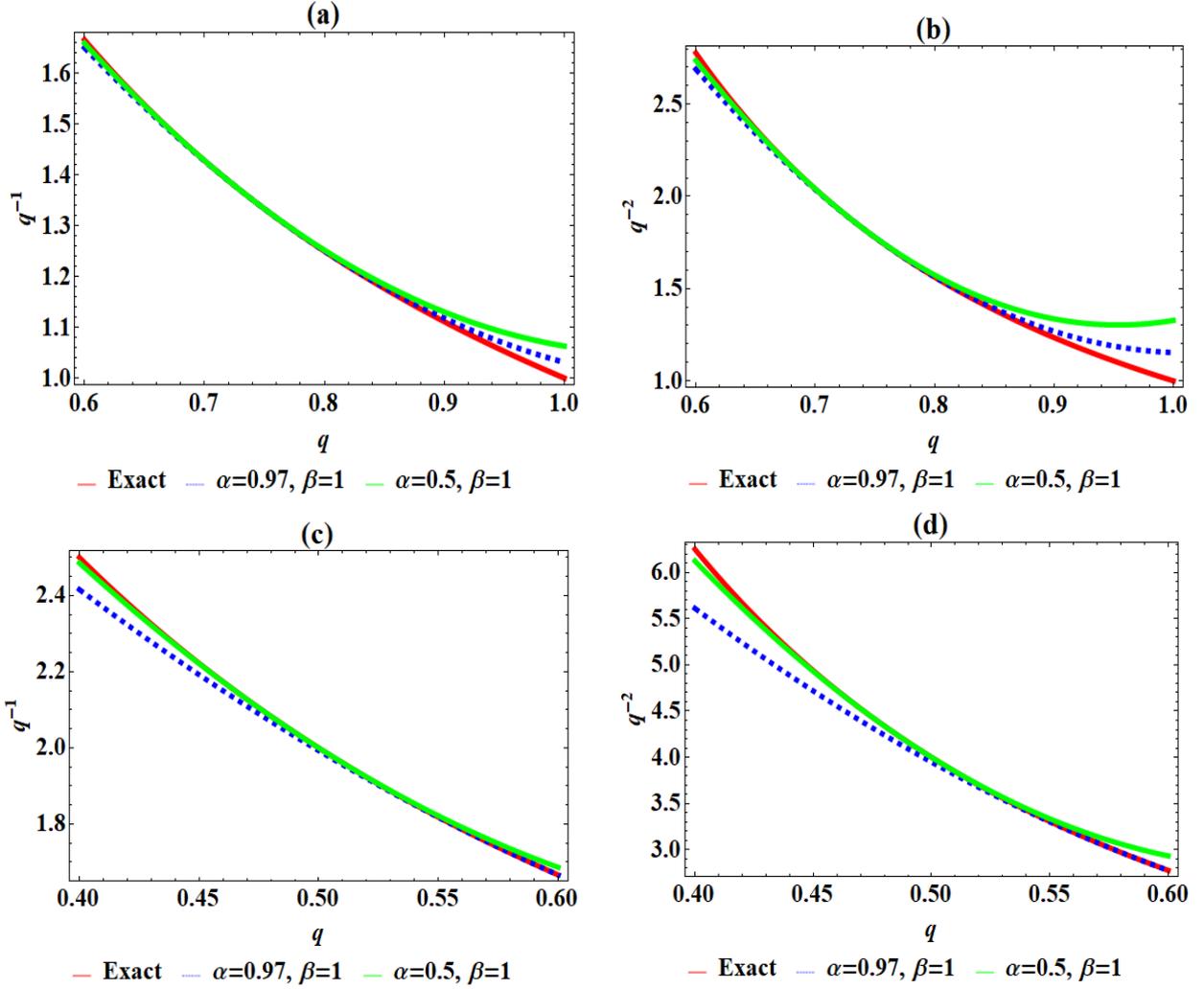

Figure 2. Pekeris approximation around $q = \delta$. $(a, b)$ Bottomonium. $(c, d)$ Charmonium

## 5-Conclusion

The bound state solutions of the fractional SE under a spin-dependent Cornell potential which satisfied the confinement force and Columbus force. By using the convectional Nikiforov-Uvarove method, we obtained the energy eigenvalues and associated wave functions in the fractional forms. At $\alpha = \beta = 1$, we obtained the results of Ref. [52] which represents a special case from the present results.

The energy spectra were applied to obtain the masses of the heavy mesons such as the bottomonium, charmonium and bottom charm. The obtained masses were found to be in good agreement with existing works and available experimental data. The results indicate that the fractional Schrodinger equation improves the mass spectra and the presence of the fractional parameters accounts for the short-range gluon exchange between the quark-antiquark interactions and the linear confinement phenomena which is associated with the quantum chromo-dynamic and phenomenological inspired potential models in particle and high-energy physics. We hope to extend this work to extreme conditions as in Refs. [56, 57].